\PassOptionsToPackage{table,xcdraw}{xcolor}
\documentclass[manuscript,screen]{acmart}
\usepackage{graphicx}
\usepackage{todonotes}
\usepackage{listings}
\definecolor{dkgreen}{rgb}{0,0.6,0}
\definecolor{dred}{rgb}{0.545,0,0}
\definecolor{dblue}{rgb}{0,0,0.545}
\definecolor{lgrey}{rgb}{0.9,0.9,0.9}
\definecolor{gray}{rgb}{0.4,0.4,0.4}
\definecolor{darkblue}{rgb}{0.0,0.0,0.6}
\lstdefinelanguage{cpp}{
      basicstyle=\small \ttfamily \color{black},   
      breakatwhitespace=false,       
      breaklines=true,               
      captionpos=b,                   
      commentstyle=\color{dkgreen},   
      deletekeywords={...},          
      escapeinside={\%*}{*)},                  
      language=C++,                
      keywordstyle=\color{purple},  
      morekeywords={BRIEFDescriptorConfig,string,TiXmlNode,DetectorDescriptorConfigContainer,istringstream,cerr,exit}, 
      identifierstyle=\color{black},
      stringstyle=\color{blue},      
      rulecolor=\color{black},        
      showspaces=false,               
      showstringspaces=false,        
      showtabs=false,                
      stepnumber=1,                   
      tabsize=5,                     
      title=\lstname,                 
    }

\usepackage{amsmath}
\usepackage{amsfonts}
\usepackage[linesnumbered,ruled]{algorithm2e}
\usepackage{enumitem}
\newcolumntype{Y}{>{\centering\arraybackslash}X}
\usepackage{subcaption}
\usepackage{stfloats}
\usepackage{float}
\usepackage{tabularx}
\makeatletter
\newcommand{\removelatexerror}{\let\@latex@error\@gobble}
\makeatother

\usepackage{layouts}

\AtBeginDocument{%
  }

\setcopyright{acmlicensed}
\copyrightyear{2024}
\acmYear{2024}
\acmDOI{XXXXXXX.XXXXXXX}


\newcommand{\Intel}{}

\newcommand{\IntelOrthe}{Intel }

\makeatletter
\renewcommand\@oddhead{\thepage}
\makeatother

\author{Mathys Jam}
\email{mathys.jam@uvsq.fr}
\affiliation{%
    \institution{Université Paris-Saclay, UVSQ, LI-PaRAD}
    \country{France}
    \city{Guyancourt}
}
\author{Eric Petit}
\email{eric.petit@intel.com}
\affiliation{%
    \institution{Intel Corp.}
    \state{Oregon}
    \country{USA}
    \city{Hillsboro}
}
\author{Pablo de Oliveira Castro}
\email{pablo.oliveira@uvsq.fr}
\affiliation{%
    \institution{Université Paris-Saclay, UVSQ, LI-PaRAD}
    \country{France}
    \city{Guyancourt}
}
\author{David Defour}
\email{david.defour@univ-prerp.fr}
\affiliation{%
    \institution{Université de Perpignan via Domitia, UPVD, LAMPS}
    \country{France}
    \city{Perpignan}
}

\author{Greg Henry}
\email{greg.henry@intel.com}
\affiliation{%
    \institution{Intel Corp.}
    \state{Oregon}
    \country{USA}
    \city{Hillsboro}
}
\author{William Jalby}
\email{william.jalby@uvsq.fr}
\affiliation{%
    \institution{Université Paris-Saclay, UVSQ, LI-PaRAD}
    \country{France}
    \city{Guyancourt}
}
\title{MLKAPS: Machine Learning and Adaptive Sampling for HPC Kernel Auto-tuning}
\date{\today}

\begin{document}

\begin{abstract}
    Many High-Performance Computing (HPC) libraries rely on decision trees to select the best kernel hyperparameters at runtime, depending on the input and environment.
    However, finding optimized configurations for each input and environment is challenging and requires significant manual effort and computational resources.
    This paper presents MLKAPS, a tool that automates this task using machine learning and adaptive sampling techniques.
    MLKAPS generates decision trees that tune HPC kernels' design parameters to achieve efficient performance for any user input.
    MLKAPS scales to large input and design spaces, outperforming similar state-of-the-art auto-tuning tools in tuning time and mean speedup. We demonstrate the benefits of MLKAPS on the highly optimized \IntelOrthe MKL \texttt{dgetrf} LU kernel and show that MLKAPS finds blindspots in the manual tuning of HPC experts. It improves over $85\%$ of the inputs with a geomean speedup of $\times1.30$.
    On the \IntelOrthe MKL \texttt{dgeqrf} QR kernel, MLKAPS improves performance on $85\%$ of the inputs with a geomean speedup of $\times1.18$.
\end{abstract}
\maketitle

\section{Introduction}

High-Performance Computing (HPC) engineers have adopted strategies to ensure that kernels perform well in different environments by configuring and
dynamically selecting the best implementation at runtime, depending on the user's input and
execution context. Software that uses decision trees~\cite{decision_trees_overview}, such as the Intel Math Kernel Library (\Intel MKL)\cite{mkl_doc}, uses an internal set of predefined optimized configurations of internal parameters. Indeed, kernels rely on two sets of runtime parameters: input parameters describing the data and arguments from the user, and design parameters selecting different algorithmic versions or runtime behavior. We aim to efficiently explore and generate the selection mechanism that associates a design configuration with any given input parameters. Increasing the number of available configurations and the complexity of the selection mechanism improves the performance and adaptability of the kernel at the expense of runtime decision cost and represents a complex trade-off for optimization. 

Furthermore, the curse of dimensionality~\cite{curse_of_dimensionality} makes it exponentially harder to find good design configurations as the size of the input and the design space increases. Recent hardware architectures with many cores, different vector sizes, numa effects, and the use of GPU accelerators contribute to larger design spaces.
Additionally, solutions relying on manual exploration are subject to human bias and lead to blind spots (regions of the input space where the configuration is sub-optimal due to an unexplored combination of parameters). As an exhaustive search becomes intractable for large design spaces, alternative solutions must be found for efficient exploration. In the \texttt{dgetrf} kernel (LU) from \IntelOrthe MKL, the design space contains \(4.6 \times 10^{13}\) possible internal configurations and a target region of interest of \(1.6 \times 10^7\) possible inputs (more details in Sec. ~\ref{sec:improving_upon_the_mkl}). This is a key limitation for the scalability of sampling based auto-tuning methods that we address in this work. To address this, we explore a new adaptive sampling heuristic.

 This article describes MLKAPS (Machine Learning for Kernel Accuracy and Performance Studies), an open-source auto-tuning framework targeting high-performance software design problems. MLKAPS automatically generates decision trees to map every input to optimized design parameters at runtime. The proposed methodology is hardware-agnostic and does not make assumptions about the optimization objectives. The former can be the kernel's execution time, numerical accuracy, or energy, depending on the user's needs. In this paper, though, we are focusing on execution time. The tool is released under an open-source license at
\href{https://github.com/MLCGO/MLKAPS}{https://github.com/MLCGO/MLKAPS}. 
 This tool can benefit many high-performance libraries, as we demonstrate on three kernels from the \IntelOrthe Math Kernel Library and ScaLAPACK.

Our contributions and key results are the following:

\begin{itemize}
    \item The MLKAPS framework, an integrated workflow to automatically generate decision trees to select optimized design parameters for any given set of inputs. Its only inputs are a description of the parameters and a kernel to evaluate configurations. 
    
    \item  We describe an implementation of statistical adaptive sampling (based on space-filling designs and variance-based techniques~\cite{paper_ask}) and a new sampling strategy we developed, GA-Adaptive, to overcome the high-dimensional space exploration problem. We demonstrate that GA-Adaptive outperforms other competitive %
    sampling strategies for kernel optimization on large input regions.

    \item  We demonstrate the benefits of the decision trees generated by MLKAPS on two different hardware architectures, using high-performance kernels from \IntelOrthe  MKL: \texttt{dgetrf} (LU) and \texttt{dgeqrf} (QR). Those kernels were highly tuned by the MKL team, and offer complex objective spaces representative of real industrial applications. 

    \item While sampling a very small fraction ($\approx \frac{1}{10^{10}}$) of the total design space, MLKAPS improves \IntelOrthe MKL \texttt{dgetrf} performance on \(85\%\) of the inputs by an geometric mean speedup of \(30\%\). The mean regression (slowdown) is $0.90$ and $1.46$ for progressions (speedup > 1.0). Furthermore, the quality of MLKAPS results continuously improves with the number of samples, and the resulting design configurations and speedup are not the same for the two architectures used, showcasing that MLKAPS adapts to its environment.

    \item We compare MLKAPS with the well-known Optuna~\cite{optuna_2019} autotuner on the \IntelOrthe MKL \texttt{dgeqrf} (QR) kernel. MLKAPS obtains a geomean speedup of $36\%$ over Optuna.

    \item We compare MLKAPS to GPTune~\cite{gptune}, a state-of-the-art autotuner, and show that MLKAPS has better performance on their ScaLAPACK use-case and our case of \IntelOrthe MKL \texttt{dgetrf} (LU).  The MLKAPS pipeline also scales to a large sample count with relatively low overhead compared to GPTune.
\end{itemize}

This paper is organized as follows: In Section~\ref{sec:problem_definition}, we define the optimization problem in large dimensional space that MLKAPS aims at solving. In Section~\ref{sec:related_work}, we review state-of-the-art auto-tuning approaches and discuss how they compare to MLKAPS. In Section~\ref{sec:mlkaps_description}, we detail MLKAPS
design principles, pipeline, and the adaptive sampling methods we explored to improve the scalability and final optimization results.
In Section~\ref{sec:results}, we demonstrate the scalability and tuning benefits of using MLKAPS on highly hand-optimized versions of HPC kernels.
Sections~\ref{sec:limitations} and~\ref{sec:conclusion} conclude and give insight into future works.

\section{Problem definition} \label{sec:problem_definition}

To illustrate the 
kind
of tuning problems we address, let us consider a naive \texttt{OpenMP}~\cite{openmp21} kernel, which sums every element of an input matrix (Figure \ref{lst:simple_kernel_example}).
The execution of this kernel is affected by various parameters, which can be classified as input and design parameters.
In practice, such parameters can be of different types, such as real, integer, categorical, or boolean.
As illustrated in Figure \ref{lst:simple_kernel_embeded_trees}, the MLKAPS objective is to produce a decision tree that selects the best design parameters (\texttt{T}) according to the input parameters (\texttt{n}, \texttt{m}). The decision trees are generated as C code to be embedded and shipped with the kernel for predictions at runtime.

\begin{figure}
    \begin{minipage}{.5\textwidth}
        \begin{lstlisting}[language=cpp]
  double sum(double* matrix, 
           int n, int m, int T) {
    double sum = 0;
    #pragma omp parallel for \ 
    reduce(+: sum) num_threads(T)
    for (int i = 0; i < n*m; i++)
      sum += matrix[I];
    return sum;
  }
        \end{lstlisting}
        \caption{\textbf{Illustrative kernel summing the elements of a matrix} with three input parameters (\texttt{matrix}, \texttt{n}, \texttt{m}), and one design parameter (\texttt{T}).}
        \label{lst:simple_kernel_example}
        \Description{We can define an illustrative kernel, a C++ function that sums every element of a matrix using OpenMP. This C++ function takes four arguments, n and m, the size of the matrix, a pointer to the data, and T, the number of threads}
    \end{minipage}\hspace{.25cm}
    \begin{minipage}{.45\textwidth}
    \vspace{.1cm}
        \begin{lstlisting}[language=cpp]
  double sum_tuned(double* matrix, 
           int n, int m) {
      // T (number of threads) 
      // is tuned depending
      // on the input parameters
      int T = decision_tree(n, m);
      return sum(matrix, n, m, T);
  }
        \end{lstlisting}
        \caption{\textbf{MLKAPS generates a tuning decision tree} to select the best number-of-threads \texttt{T} for a given input context.}
        \label{lst:simple_kernel_embeded_trees}
        \Description{We can now create a new version of this kernel, that automatically computes the optimal number of threads T using a decision tree. The decision tree takes the matrix size, n and m, as inputs. This value is then forwarded to the original kernel. As such, the user doesn't have to worry about T when calling the kernel.}
    \end{minipage}
\end{figure}

{\bf Input parameters} correspond to the task specification and cannot be tuned. They are either defined by the user (e.g., \texttt{matrix} values, \texttt{n}, and \texttt{m}) or correspond to fixed parameters of the execution environment (e.g., available memory, available number of cores.)

{\bf Design parameters} refer to the set of parameters MLKAPS optimizes. Unique combinations of design parameter values are referred to as \emph{configurations}.
These parameters may be internal to the kernel (e.g. algorithmic variant, number \texttt{T} of threads in the previous example) or the execution environment (e.g., thread placement, core frequency, compiler flags.)

{\bf The decision tree} provides optimized values while minimizing execution time or any other objective defined by the user. This decision tree is evaluated at runtime to select the best configuration depending on the input(s).

\section{Related works} \label{sec:related_work}

In \cite{autotuning_survey}, the authors survey the field of auto-tuning programs
and classify the different frameworks based on their approach used for each different component of the tuning problem they are addressing:

\begin{enumerate}[label=(\Alph*)]
    \item Tuning rating method (how are solutions evaluated): empirical evaluation or model-driven.
    \item Tuning time (when does the tuning occur): online tuning, offline tuning
    \item Search space exploration algorithm: (biased) random search, brute-force,
          parallel search, profile-based optimization, and pruned search space tuning
    \item Tuning scope: program-level, section-level, procedure-level
\end{enumerate}

Since MLKAPS operates in a black-box setup, it is kernel-agnostic and can target
all three tuning scope levels mentioned in (D). In the following, we review the
related works using the criteria (A), (B), and (C).

\subsection{Tuning rating method}

Tuning rating is the method used to evaluate a configuration and how it compares to others.
On one hand, Atlas~\cite{atlas},
Spiral~\cite{spiral_code_generation_for_dsp_transforms}, and
FFTW~\cite{a_fast_fourier_transform_compiler} are auto-tuned libraries that use
empirical rating: the configurations are executed in the target environment, which is costly but also brings the most information. More recently, Bolt~\cite{bolt_tensor_tuner} and AutoTVM~\cite{AutoTVM} also use empirical evaluation to evaluate a set of candidate configurations for Deep Neural Network optimization.

On the other hand, Bergstra et al.~\cite{machine_learning_for_auto_tuning_with_boosted_regression_trees} implement
a model-driven strategy: an approximation of the kernel's objectives is used to
rank the tuning solutions. This surrogate model can be built in two ways: using
heuristics/models based on expert knowledge of the kernel or ML models
trained on the kernel. Using a model allows the optimization process to rate as many configurations as needed. Hand-tuned cost models are error-prone and subject to bias but are relatively cheap to produce. AutoTVM and Ansor~\cite{Ansor} build Gradient Boosting Decision Trees (GBDT)~\cite{Benchmarking_and_Optimization_of_Gradient_Boosting_Decision_Tree_Algorithms}, a machine learning ensemble method based on decision trees, to perform a preliminary low-cost evaluation of the solutions.

\subsection{Tuning time}

In Atlas~\cite{atlas} or Patus~\cite{PATUS}, the tuning process occurs offline before the user's program is run. By reloading the optimized solutions, runtime overhead is minimal. However, offline tuning can suffer from a lack of contextual information available at runtime.

In online tuning, the optimization process occurs at runtime
with contextual knowledge. In \cite{input_aware_auto_tuning_of_compute_bound_hpc_kernels}, hill-climbing is
used at runtime in a model-driven approach. In \cite{socrates}, multiple tools
monitor and adapt the kernel tuning to honor specific application
requirements. The difficulty in this approach is the trade-off between
optimization overhead and the quality of the tuning results: the optimization
cost must not overshadow the benefits realized in application performance.

In Bolt~\cite{bolt_tensor_tuner}, domain-specific knowledge, hardware information, and heuristics are used to reduce the tuning time to around 20 minutes per tuning iteration. This is a compromise between black box tuners, which can run for days, and hand-tuned cost model approaches. This is particularly useful for their Deep Neural Network target applications, as users can try different neural network layouts in quick iterations. Felix~\cite{Felix} proposes a similar approach, using pre-trained models as predictor using the Tenset~\cite{Tenset}. They evaluate models such as MLP and boosting trees to predict kernels' performance based on their features (number of load, store, reuse distance\dots) and hardware features (cache sizes\dots) to select candidate loop scheduling in tensor computation. While using a predictor handling large range of input characteristics, the prediction and final exploration are too expensive for runtime evaluation, and requires kernel bounds to be known at compile time. This perfectly fit the JIT use case they aim at, but cannot cover the requirement of HPC applications such as math library. However, this could be used to tune lower level kernels used in the math libraries algorithm. 

\subsection{Exploration algorithm}
The last criterion is the algorithm used to explore the search space. Atlas and
many other auto-tuned libraries rely on brute-force search with an empirical
rating. Adaptive sampling strategies fall in the biased random search category,
which covers a wide range of techniques.
They typically start with a random walk in the search space before introducing a bias towards the most promising solutions.
Optuna~\cite{optuna_2019} uses a combination of Covariance Matrix Adaptation Evolution Strategy (CMA-ES)~\cite{tutorial_cmaes} and Tree-structured Parzen Estimator (TPE)~\cite{tree_parzen_estimator} to explore the design space, using empirical evaluations paired with an early-stopping criterion that stops trials that perform worse than the best-known solution. We compare MLKAPS and Optuna in Section ~\ref{sec:mlkaps_vs_optuna}.

Luszczek et al.~\cite{combining_multitask_optimization} uses a Bayesian optimization technique combined with multi-task and transfer learning to solve
an auto-tuning problem. GPTune~\cite{gptune} builds Gaussian-Process (GP) models in a Bayesian optimization pipeline. However, while most work presented here only outputs a unique configuration for the application, GPTune works on a problem similar to MLKAPS, mapping multiple inputs to dedicated configurations.
Furthermore, while MLKAPS uses decision trees to generalize, GPTune implements a strategy called TLAII, which can infer configurations for new inputs without resampling, based on previous results.
We compare MLKAPS and GPTune in Section \ref{sec:comparison_to_bayesian}.

In Bolt~\cite{bolt_tensor_tuner}, and AutoTVM~\cite{AutoTVM} authors use domain-specific knowledge to inform their search. Bolt uses hardware information to restrain the space of potential configurations based on a set of rules for Nvidia GPUs. AutoTVM uses an iterative algorithm based on GBDT~\cite{Benchmarking_and_Optimization_of_Gradient_Boosting_Decision_Tree_Algorithms} and simulated annealing~\cite{simulated_annealing}.

\section{MLKAPS description} \label{sec:mlkaps_description}
\begin{figure}
    \centering
    \includegraphics[width=\linewidth]{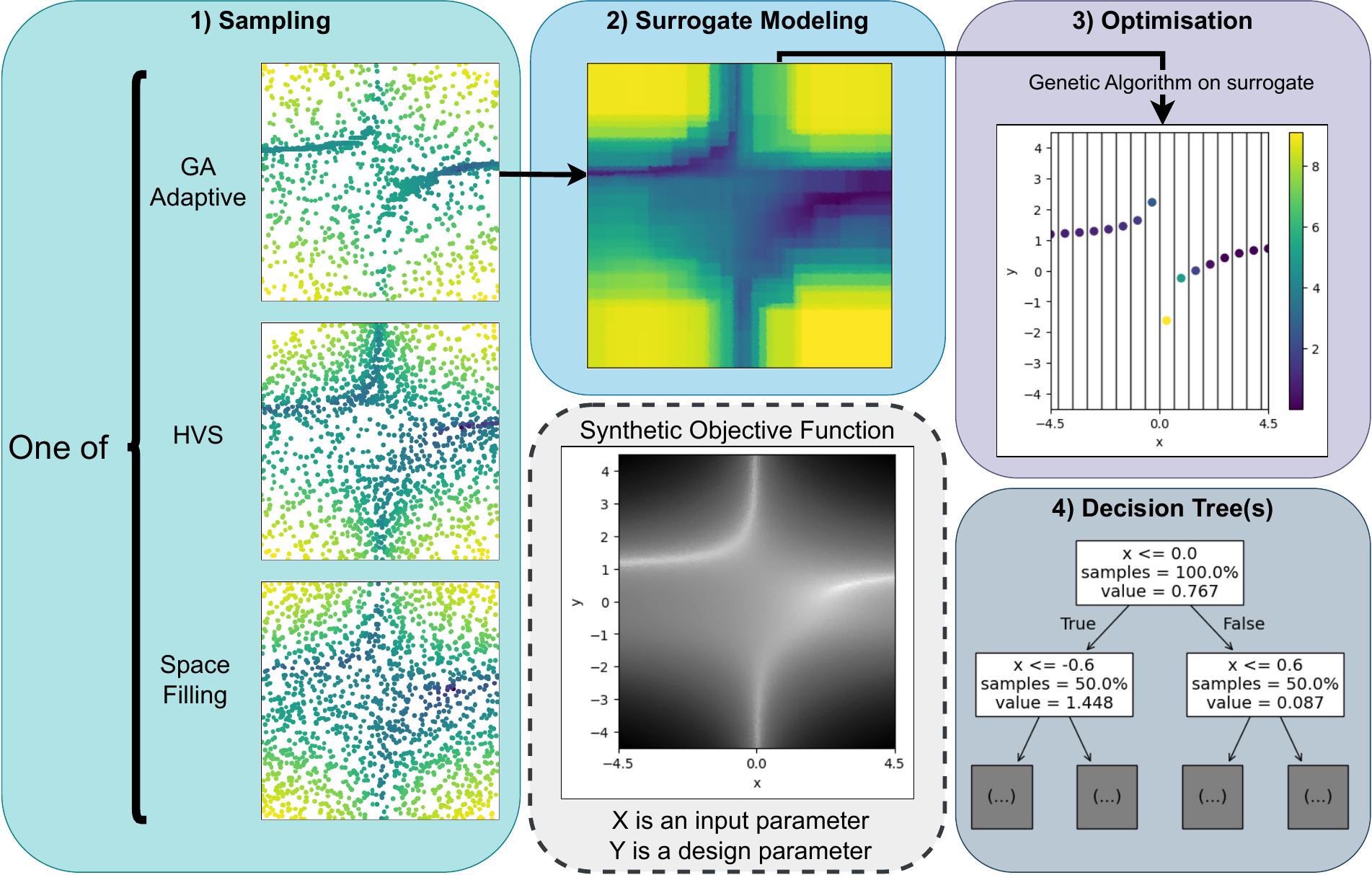}
    \caption{\textbf{Flowchart of the MLKAPS auto-tuning pipeline.} This example illustrates the different steps to build a decision tree that maps one design parameter $Y$ to one input parameter $X$. For an industrial application with a much larger number of parameters, it would be intractable to map the objective function exhaustively as shown here.}
    \label{fig:example_dt_building_process}
    \Description{A flowchart showcasing MLKAPS pipeline. A 2D function is plotted as a map, showcasing the optimal Y value for every possible X. First, three scatter plots are displayed, showcasing where GA-Adaptive, HVS, and random, three sampling algorithms, are measuring the objective function. We observe that GA-Adaptive mostly measures around the optimal Y value. HVS measurements are more spread out focusing on high variance region, leaving others with  lower  density of samples. Finally, the random sampling uniformly distributes measurements over X and Y. Second, a 2D map shows the predicted objective using a model fitted using GA-Adaptive measurements. The map is very accurate around the optimal Y value, but has very poor accuracy everywhere else. Third, another 2D map, with X axis split in multiple bins column columns. In every column, a colored dot appears where the optimizer picks an optimal Y solution. Overall, the dots follow the pattern of the optimal values from the true objective function. Fourth and last, a simple decision tree is shown, selecting a Y for each bins of X.}
\end{figure}

MLKAPS (Machine Learning for Kernel Accuracy and Performance Studies) uses a model-driven rating method powered by GBDT models from the LightGBM~\cite{LightGBM} Python library trained using empirical evaluations. It was designed to perform equally well on all regions of the input space, leverage transfer learning to increase efficiency by sharing knowledge across regions, and scale to large design spaces. Figure~\ref{fig:example_dt_building_process}
shows the pipeline implemented in MLKAPS.

We divide MLKAPS' pipeline into two components: first, sampling and modeling, detailed in Section~\ref{sec:sampling_and_modeling}, where we collect knowledge about the kernel and refine the GBDT model; and second, optimization and decision tree creation, detailed in Section~\ref{sec:optimizing}, where we use the surrogate model to find a set of optimized configurations.

\subsection{Sampling and Modeling}
\label{sec:sampling_and_modeling}
Let us assume that we are initially given a black-box kernel that measures the target objective for any given inputs and design parameters. We aim to gather a dataset to train a machine-learning model to infer the objective value. We will refer to a kernel execution as a sample. Finally, we will use this as a surrogate model for configuration rating, allowing us to replace costly kernel sampling with the low-cost inference of the model. 

As measuring every possible configuration is intractable due to the combinatorial explosion and sample execution cost, the measured samples must be carefully selected. We aim to maximize how much useful information the surrogate model can extract from each sample towards our optimization objective. We implemented various adaptive sampling approaches, trading between exploring new input areas and exploiting current knowledge to determine the next samples of interest. MLKAPS implements three main strategies described in the next sections: space-filling sampling, variance-based sampling, and optimization-driven sampling. Those strategies are illustrated in Figure \ref{fig:example_dt_building_process}, and will be compared in Section~\ref{sec:results}.

\subsubsection{Space-filling sampling}
is the simplest method for sampling points spread over the entire input space.
It aims to bias for spatial coverage, as shown in figure~\ref{fig:example_dt_building_process}, by ensuring that all configurations have an equal probability of being sampled. A simple example is random sampling.
Latin Hypercube Sampling (LHS)~\cite{Latin_hypercube_Sampling} is an
alternative sampling method that divides each parameter into intervals and
ensures that each interval is sampled exactly once. This approach provides
better coverage of each of the parameter space and captures variations and interactions
more effectively, which can help reduce uncertainty arising from measurement noises.

\subsubsection{HVS sampling}
is another strategy to characterize a function
efficiently by biasing the sampling based on variance. In \cite{paper_ask}, authors propose an iterative sampling
strategy based on a partitioning of the parameter space, named Hierarchical Variance Sampling (HVS). LHS is used for bootstrapping the method. The samples are then partitioned using a decision tree, and the partition size and variance estimator are computed. The next iteration of sample selection makes a trade-off between exploiting current variance estimator information and exploring new regions, therefore, the samples are distributed among the partitions depending on the product
$\text{size} \times \text{variance}$. 
  In figure~\ref{fig:example_dt_building_process}, some regions have a very low sample density due to their low variance, which is better redistributed to characterize complex regions of the objective function.

The variant HVS-relative (HVSr)~\cite{paper_ask} uses an estimator of the coefficient of variation for cases where the objective function has a large range of non-uniformly distributed values. 

Furthermore, some ill-configurations can give really poor execution time and introduce high-variance regions in the objective space. As a consequence, HVS can spend a huge portion of its sampling budget trying to capture such configurations, while not providing useful information on the potential local optima. The method does not distinguish between variance introduced by good and bad configurations. We introduced an upper bound in the objective function to prevent HVS from overly-sampling outlier configurations. This proved very effective in the case of \texttt{dgetrf} (LU) and \texttt{dgeqrf} (QR), with negligible quantities of wasted samples in our final set of experiments.

\subsubsection{GA-adaptive sampling}
We develop a new optimization-driven sampling strategy, GA-adaptive, with the rationale that the surrogate model does not need to learn the entire objective space; it should trade off generalization for high accuracy in regions containing good configurations. This is visible in figure~\ref{fig:example_dt_building_process}, where almost no points are chosen outside the optimal neighborhoods, trading global accuracy for better predictions on optimized configurations.

It deals with the exploration/exploitation dilemma~\cite{exploration_exploitation_dilemma} by replicating the MLKAPS' optimization phase using an $\epsilon$-decreasing strategy~\cite{epsilon_decreasing_bianchi}.

\begin{figure}
    \removelatexerror
    \begin{algorithm*}[H]
        \SetKwInOut{Input}{input}\SetKwInOut{Output}{output}
        \Input{$b$ ratio of initial bootstrap samples \newline
            $i$ the initial ratio of points taken with the GA sampler \newline
            $f$ the final ratio of points taken with the GA sampler \newline
            $s$ the number of samples to take per iteration\newline
            $n$ the total number of samples\newline}
        \Output{set of sampled points}

        $Samples \leftarrow$ \textbf{BootstrapLHS}($b\times n$)
        \tcp*{\footnotesize{Bootstrap the algorithm with $b \times n$ LHS samples}}
        \While{$|Samples| < n$}{
        $p \leftarrow |Samples| / n$\;
        $\epsilon \leftarrow  i + (f - i) \times p$\;
        $Model \leftarrow \textbf{GBDT}(Samples[Parameters], Samples[Objective])$
        \tcp*{\footnotesize{Fit a surrogate model}}
        $OptimPoints \leftarrow \textbf{PickRandomInputs}(\epsilon \times s)$ \tcp*{\footnotesize{Select random inputs for GA}}
        $New_{ga} \leftarrow  \textbf{GA}(OptimPoints, Model)$
        \tcp*{\footnotesize{Exploitation: Optimize with GA on $\epsilon \times s$ new inputs}}
        $New_{sub} \leftarrow \textbf{SubSampler}((1 - \epsilon) \times s, Samples)$ \tcp*{\footnotesize{Exploration: Pick leftover samples with sub-sampler}}
            $Samples \leftarrow Samples \cup New_{sub} \cup New_{ga}$\;
            }
            return $Samples$\;
    \end{algorithm*}
    \caption{\textbf{GA-Adaptive core loop}. This algorithm uses a strategy similar to epsilon-decreasing~\cite{epsilon_decreasing_bianchi} to solve the exploration-exploitation dilemma by linearly increasing the number of samples taken for exploitation. Note how the sampling-modeling-optimization steps are similar to the global pipeline. At each iteration, the percentage of points taken with GA and the sub-sampler is computed as a linear interpolation between the initial and final ratios, controlled by the completion percentage (i.e., at $50\%$ completion with an initial ratio of $0$ and a final ratio of $0.8$, GA will pick $40\%$ of points in the next iteration).
    }\label{alg:ga_adaptive}
    \Description{A pseudo-code for the GA-Adaptive algorithm. This algorithm takes 5 inputs: B, the fraction of samples used for bootstrapping. i, the initial fraction of points taken with G.A. f, the final fraction of points taken with G.A. s, the number of samplers per iteration. And n, the total number of samples.
        
    The algorithm first bootstraps according to the bootstrapping fraction and the total number of samples. It then iterates until all samples have been collected.
    
    At each iteration, we first compute the percentage of completion. We then compute the fraction of points taken with GA, by interpolating between the initial fraction, i, and the final fraction, f, according to the completion percentage. We then fit a surrogate model based on the existing samples. We randomly choose optimization points by multiplying the fraction of points taken with GA, with the number of samples taken at each iteration. We then run genetic algorithms on every optimization points using the surrogate model. The remainder of the points are taken with the sub-sampler. Finally, the new points are merged with the existing samples for the next iteration.}
\end{figure}

Algorithm \ref{alg:ga_adaptive} shows a pseudo-code of GA-Adaptive. The
algorithm is initialized (Line 1) using LHS to take the initial points and gain enough
knowledge to build a surrogate for the GA sampler.

During each sampling iteration, a surrogate model is fitted on past samples. Input configurations are randomly selected and optimized using a Genetic Algorithm on the surrogate model. The more the algorithm samples, the more accurate the model will be on good configurations. Therefore, the GA runs will likely converge towards the optima thanks to the following effects:
\begin{itemize}
    \item If the model is overly optimistic on a configuration, it will be selected for sampling and the true value added to the dataset, correcting the model so that future iterations will avoid it. This also applies to the final optimization phase in MLKAPS' pipeline, as it uses the same modeling and optimization algorithms.
    \item If the model prediction was right, the local accuracy of the surrogates around promising solutions is improved. This allows us to differentiate between two solutions with similar performance, reducing uncertainty caused by noise.
\end{itemize}

In the first iterations, exploration is favored at the expense of exploitation to consolidate the surrogate model. Then, as the model accuracy improves over time, we move towards exploitation and focus on improving knowledge of good configurations. This proportion follows a linear progression between user-defined starting and end bounds according to the iteration count.

GA-Adaptive behavior depends on a set of hyper-parameters: the algorithmic choice for the optimization process (GAs in our implementation), the bootstrapping algorithm (LHS), the sub-sampler (HVSr by default), and the ratio of points taken with GA ($\epsilon$), usually starting at 0.0 with a final value of 1.0 at the last iteration. 

\subsubsection{Modeling} is the last step of the first stage, where we train the main surrogate model using the previously sampled dataset.
we focused on gradient-boosting decision trees (GBDT)~\cite{Benchmarking_and_Optimization_of_Gradient_Boosting_Decision_Tree_Algorithms} from \texttt{LightGBM}~\cite{LightGBM}. It allows efficient handling of all variable types, including categorical, and offers scalable and accurate modeling. GBDT is used by many existing autotuners~\cite{AutoTVM, machine_learning_for_auto_tuning_with_boosted_regression_trees} 

Plots~\ref{fig:global_accuracy_mlkaps} and~\ref{fig:local_accuracy_mlkaps} show different model metrics have different convergence behavior, making it challenging to select the most suitable metric for our optimization objective. Since we have no prior knowledge of the objective space, we usually select Mean Absolute Error (MAE) for the model's metric, which works well in the cases studied in this paper. In cases where the objective value can cover a huge range of values, Mean Average Percentage Error (MAPE) was observed to improve the tuning results significantly. However, the only reliable indicator of model performance we have available is comparing speedup or mean performance using the solutions the optimizer found on the surrogate. Those metrics were the most helpful in our development of the MLKAPS pipeline.

We used two approaches to tune the surrogate's hyperparameters: manual tuning and Optuna~\cite{optuna_2019}. Using Optuna with an MAE metric led to worse optimization than a hand-tuned model. We believe this is due to the need for a dedicated metric for the model, one that measures the accuracy on the best configurations. Given the impact of surrogate quality on our result, we are considering defining and experimenting with such new metrics in future work.

\subsection{Optimization and Decision trees}
\label{sec:optimizing}

This step computes a set of optimized configurations across the input space using GAs. We use \texttt{pymoo}'s~[22]\cite{pymoo} implementation of NSGAII~\cite{paper_nsga2} for its robustness and generalization capabilities. We first select "optimization points" by building a regular grid over the input space, with a user-defined size in each dimension. We then run one GA instance per point in the grid and find the best configuration, one that leverages the local behavior of the objective function on a given input region. 
Optimal performance in HPC usually occurs on cliffs~\cite{performance_cliffs},
dictated by cache size, pipeline saturation, or other discrete thresholds.
As such, the optimization quality is significantly impacted by the coarseness of the grid. A coarse grid might lead to suboptimal choices, while a finer grid better leverages locality. There is a trade-off between the size of the grid and the time budget for each point.
Running the optimization phase on a surrogate model with cheap predictions allows us to increase the density of the grid. And since MLKAPS sampling is not restricted by the grid, the grid point configurations will be informed by their local neighborhood and not limited by strictly local information, resulting in a better generalization of the solution and limiting potential over-fitting. In practice, we did not observe improvement by selecting 24x24 grid compared to 16x16 and as a consequence select the later as default value for our experiment unless specified otherwise.

The final step of the pipeline is to build one or more decision trees on the previous grid. For this, we use \texttt{scikit-learn}'s Decision Tree Regressor for continuous variables and classifier for categorical variables. Those trees are stored in a pickled file, and generated as C code for the user to embed in his kernel. We currently build one decision tree per design parameter, whose outputs are combined to get the full configuration.

As the runtime overhead of the decision process could take away the potential gains, the application often has a limited time budget to decide the configuration to use. Deep trees maximize choice locality but can have a significant overhead; shallow trees will select a trade-off between multiple configurations, reducing the overhead at the cost of suboptimal configurations.

\section{Results} \label{sec:results}

\begin{figure}
    \centering
    \caption{\textbf{Hardware architectures used for the experiments.}}
    {\scriptsize
        \begin{tabularx}{\textwidth}{|l|XXXXXXXXX|}
            \hline
            \rowcolor[HTML]{C0C0C0}
                                                                 &
            \multicolumn{1}{l|}{\cellcolor[HTML]{C0C0C0}CPU}     &
            \multicolumn{1}{l|}{\cellcolor[HTML]{C0C0C0}freq.}   &
            \multicolumn{1}{l|}{\cellcolor[HTML]{C0C0C0}cores}   &
            \multicolumn{1}{l|}{\cellcolor[HTML]{C0C0C0}threads} &
            \multicolumn{1}{l|}{\cellcolor[HTML]{C0C0C0}L1}      &
            \multicolumn{1}{l|}{\cellcolor[HTML]{C0C0C0}L2}      &
            \multicolumn{1}{l|}{\cellcolor[HTML]{C0C0C0}L3}      &
            \multicolumn{1}{l|}{\cellcolor[HTML]{C0C0C0}Ram}     &
            Type                                                   \\ \hline
            \cellcolor[HTML]{C0C0C0}\textbf{KNM}                 &
            Intel Knights Mill                                   &
            1.5GHz                                               &
            72                                                   &
            288                                                  &
            32 KB                                                &
            36 MB                                                &
                                                                 &
            16GB                                                 &
            HBM                                                    \\ \hline
            \cellcolor[HTML]{C0C0C0}\textbf{SPR}                 &
            Intel Xeon Gold 6438M                                &
            2.2GHz                                               &
            64                                                   &
            128                                                  &
            80 KB                                                &
            2 MB                                                 &
            60MB                                                 &
            504GB                                                &
            DDR5                                                   \\ \hline
        \end{tabularx}
    }
    \label{tab:architectures}
\end{figure}

\subsubsection{Experimental Setup}

We evaluated MLKAPS on two HPC servers with widely different micro-\-architectures: an \IntelOrthe Knight Mill processor and \IntelOrthe Sapphire Rapids processor, described in Table \ref{tab:architectures} and studied three high-performance kernels: 
\begin{itemize}
    \item \texttt{dgetrf} and \texttt{dgeqrf} from  \IntelOrthe MKL. The \texttt{dgetrf} and \texttt{dgeqrf} binary Intel provided us is statically linked to a prototype derived from MKL version 2022.0 Update 2, OpenMP 4.5, and contained the kernel version specifically tuned for both the KNM and SPR. Furthermore, we explore the input and design parameters range specified by Intel for this benchmark.  
    \item \texttt{pdgeqrf} from ScaLAPACK~\cite{scalpack_user_guide}. We use ScaLAPACK Version 2.1.0 combined with OpenMPI 4.0 as provided in the GPTune Docker image\footnote{liuyangzhuan/gptune:4.5 \url{https://hub.docker.com/r/liuyangzhuan/gptune}}.
\end{itemize}

\subsubsection{\texttt{dgetrf} description}

We focus on a detailed experiment and analysis on \IntelOrthe MKL \texttt{dgetrf} kernel. This kernel performs an LU factorization based on level 3 BLAS computation. Two of its input parameters correspond to the matrix size, named $n$ and $m$, which significantly influence the execution time, as the flop count is approximately $\frac{2}{3}m^2n$ (depending on the largest dimension). We bound $1000 \leq n, m \leq 5000$ for this study according to the benchmark specification. The kernel exposes eight internal design parameters, such as the number of threads and tilling configuration, which lead to up to ten dimensions to explore during the sampling phase. When we compare to the MKL, we consider their current hand-tuned internal configurations as a reference. Recall that MLKAPS is a black-box approach: it assumes no prior knowledge about these parameters and their role, and as such, we did not reuse any information that could be extracted from this hand-tuning.
This study considers a single objective: minimizing execution time. MLKAPS uses depth $8$ decision trees trained on an optimization grid $16\times16$.

First, Section \ref{sec:sampling_algorithms_accuracy} compares the different sampling strategies and their impact on the accuracy of the GBDT model. Section~\ref{sec:sampling_algorithms_comparison} evaluates the tuning capability of each algorithm. Section~\ref{sec:improving_upon_the_mkl} demonstrates that MLKAPS improves the existing tuning of \IntelOrthe \texttt{dgetrf} on both architectures. Section~\ref{sec:comparison_to_bayesian} compares MLKAPS with the state-of-the-art auto-tuners, GPTune~\cite{gptune} and Optuna~\cite{optuna_2019}.

\subsection{Sampling strategies and model accuracy}
\label{sec:sampling_algorithms_accuracy}

\begin{figure}
    \centering
    \includegraphics{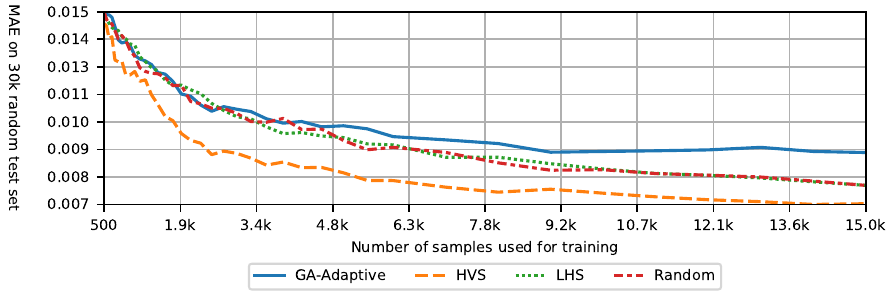}
    \caption{\textbf{(SPR): Global accuracy measured using 30k random samples with  GBDT surrogate models trained using different sampling strategies.} (lower is better). Up to 15k samples per method, same model hyperparameters. HVS outperforms the other samplers for global accuracy; thus, it is a good fit for the exploration phase.}
    \label{fig:global_accuracy_mlkaps}
    \vspace{0.5cm}
    \includegraphics{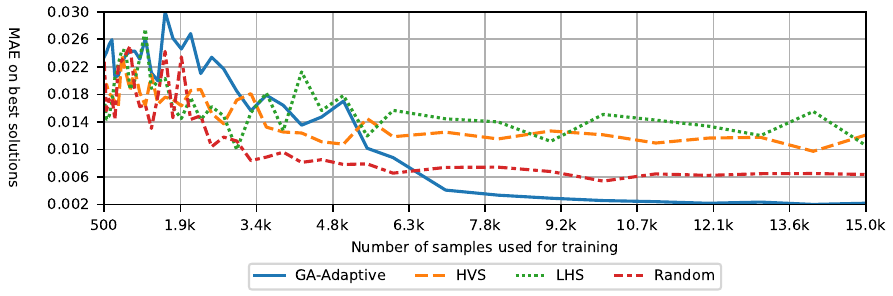}
    \caption{\textbf{(SPR): Local accuracy on the predicted best configurations (1024 per method) with GBDT surrogate models trained using different sampling strategies.} (lower is better). Up to 15k samples per method, same model hyperparameters. GA-Adaptive outperforms all methods when optimizing for local accuracy.}
    \label{fig:local_accuracy_mlkaps}

    \Description{This figure show two lineplot. The first figure shows the global accuracy, measured using Mean Absolute Error, depending on the method used to sample the training dataset of the surrogate model. HVS converges rapidly to an MAE of 0.007. LHS and Random overlap, and converge slowly to 0.008. Finally, GA-Adaptive converges quickly at first, but then stagnates to 0.01. Overall, GA-Adaptive has the worst accuracy, LHS and Random are in the middle with similar performance, and HVS beats all other methods.
    
    The second figure is similar, but it measures the error on the predicted performance of the tuning results. This local accuracy gives the model performance on the best configurations. All methods are chaotic with few samples, but GA-Adaptive converged to a mean absolute error of 0.002 after 7k samples. Random converges steadily to 0.006, while LHS and HVS converge to 0.012. Overall, GA-Adaptive outperforms the other appraoch significantly on this metric.
    }
\end{figure}

Since optimization runs on top of the surrogate model, its accuracy is critical and must be carefully evaluated for each sampling algorithm. We ran the different algorithms on \IntelOrthe MKL \texttt{dgetrf} kernel with $15k$ samples and evaluated the accuracy of the model. This experiment focuses on comparing and validating our heuristic for the GA-adaptive algorithm.

\subsubsection{Global accuracy}
First, we measured the global precision of the model using a validation data set composed of 30k random samples.
Figure \ref{fig:global_accuracy_mlkaps} shows the global precision of the resulting models. We observe that HVS outperforms the other strategies, which is expected as it was developed to maximize the accuracy gained with each sample. LHS and Random have similar performance, while GA-Adaptive has a higher error, which is expected since it sacrifices global accuracy.

\subsubsection{Local accuracy}
Then, we measured the local accuracy of each method on the output of the optimization phase with the same model. This gives us the accuracy around the predicted optimal configurations. Figure \ref{fig:local_accuracy_mlkaps} shows that GA-Adaptive has a significantly lower MAE on the best solutions, meaning it chooses the best configurations with higher confidence.

\subsection{Sampling strategies and optimization}
\label{sec:sampling_algorithms_comparison}

\begin{figure}
    \centering
    \includegraphics[width=1\textwidth]{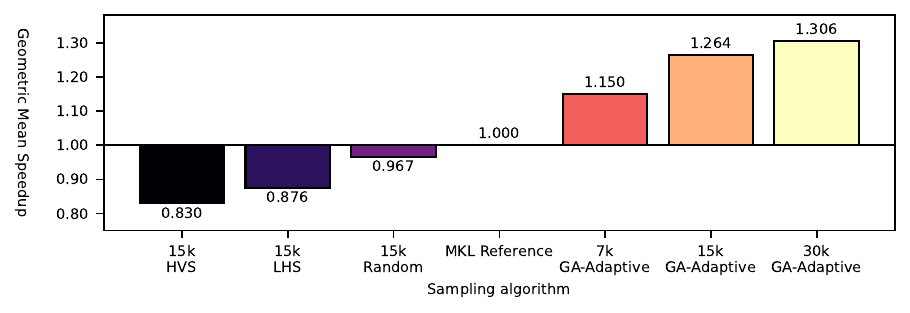}
    \caption{\textbf{(SPR): Geometric mean Speedup (higher is better)  against the MKL reference configuration on \texttt{dgetrf} (LU)}, depending on the sampling algorithm. 46x46 validation grid. 7k/15k/30k denotes the samples count. GA-Adaptive outperforms all other sampling strategies for auto-tuning. With 30k samples it achieves a mean speedup of $\times 1.3$ of the MKL dgetrf kernel.}
    \label{fig:tuning_results_ranking}
    \Description{This figure shows a barplot, centered around 1.0, reprensenting the MKL reference. Each bar corresponds to the geometric mean speedup relative to the reference, on the dgetrf (LU) kernel. In order: 15k HVS with a geomean speedup of 0.83, 15k LHS with 0.876, 15k Random with 0.967, 7k GA-Adaptive with 1.15, 15k GA-Adaptive with 1.264, and finally 30k GA-Adaptive with 1.306. Overall, only GA-Adaptive is able to beat the MKL reference. }
\end{figure}

To compare the sampling strategies implemented in MLKAPS, we built a regular grid of $46\times46$ points (for a total of $2116$ points) in the input space. We then measured the configuration predicted by MLKAPS with different sampling strategies and sample counts and computed the geometric mean speedup relative to the MKL configuration.

The result given in Fig. \ref{fig:tuning_results_ranking} and previous observations on the influence on local and global accuracy of sampling strategies endorse our heuristic that improving local accuracy using GA-Adaptive is more beneficial to tuning quality than improving the overall accuracy using HVS. It also shows that measuring the global accuracy using MAE or RMSE on random samples does not allow us to conclude on the model's quality for tuning, as GA-Adaptive performed worse than random sampling on this metric. While HVS produced the best model for global accuracy, it performs worse than random sampling for tuning this kernel.

\subsection{Improvment upon hand-tuning}
\label{sec:improving_upon_the_mkl}

In the following, we will dive deeper into how tuning solutions obtained
with MLKAPS perform when compared to the highly optimized hand-tuning of the \IntelOrthe MKL \texttt{dgetrf} kernel.

\subsubsection{Results on the KNM}
\begin{figure}[tb]
    \centering
    \includegraphics[]{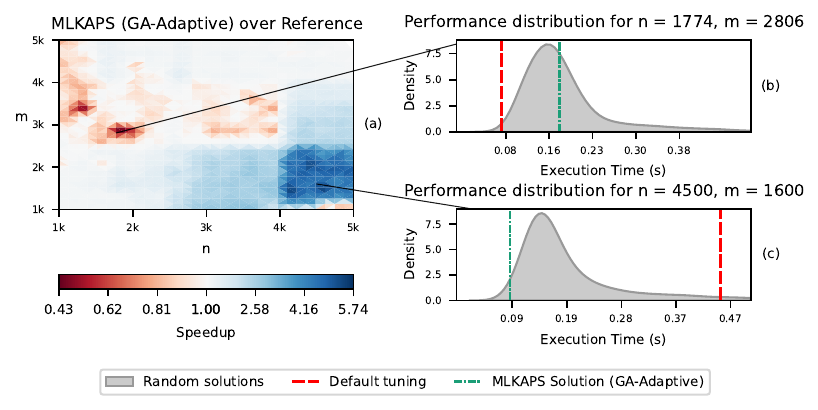}
    \caption{\textbf{(KNM): (a) Speedup map of GA-Adaptive} (7k samples) over the \IntelOrthe MKL hand-tuning for \texttt{dgetrf} (LU), higher is better. \textbf{(b) Analysis of the slowdown region} (performance regression). \textbf{(c) Analysis of the high speedup region.} $3,000$ random solutions were evaluated for each distribution.}
    \label{fig:speedup_map}

    \Description{This figures includes three plots. The first plot is a 2d heatmap showing the speedup depending on the matrix size, n and m. We observe that for m < 2300, MLKAPS shows a speedup if up to x5.74 compared to the reference MKL. The speedup increases with n. On the rest of the space, the speedup is around 1.0. Finally, we observe a few regions with significant slowdowns.
    
    The second figures is a zoom on a slowdown region. A plot shows the performance distribution of random configurations. A vertical line shows that the MKL is in the top of the distribution. A second vertical line shows that MLKAPS is near the middle of the distribution, meaning that MLKAPS is average while the MKL is very good.
    
    The last figures is a zoom on a speedup region. MLKAPS is in the top of the distribution, while the MKL is a complete outlier, in the tail of the distribution, meaning that the MKL picks one of the worst possible configurations.}
\end{figure}

Figure \ref{fig:speedup_map}~(a) shows the relative performance
of GA-Adaptive compared to the MKL tuning on the KNM architecture, using a $32\times32$ regular grid over the input space.
For this experiment, we limit the number of samples to $7000$ to keep the runtime short enough to explore MLKAPS configurations and illustrate pitfalls of not having enough samples.
We observe that MLKAPS's results are non-uniform over the input
space:
\begin{itemize}
    \item The first region, with $2500 < m < 5000$, shows an average speedup of 1,
          meaning that MLKAPS performs as well as the MKL reference. However, for some inputs, MLKAPS shows significant performance regressions.
    \item The second region, with $1000 \leq m \leq 2500$ shows significant speedups for MLKAPS, up to $\times 5$ for $n > 4000$
    \item The grid-like pattern is due to the decision trees partitioning of the input space.
\end{itemize}

To understand the speedup map, we chose one point in the best and worst regions and measured randomly chosen configurations. This stochastic sampling allows us to estimate the performance histogram in the region and see where MLKAPS and MKL choices lie in the distribution.
The first Histogram \ref{fig:speedup_map}~(b) shows the distribution of performance for  $(n=1774;m=2806)$ in a region where MLKAPS did not perform well. MLKAPS picks an average solution, while the MKL chooses among the best available. In that particular case, it was due to an incorrect prediction from the surrogate model, which is solved by increasing the number of samples until the model has converged, like shown in the previous section with figure \ref{fig:tuning_results_ranking}.

The second Histogram \ref{fig:speedup_map}~(c) shows the distribution of performance for $(n=4500;m=1600)$ in the region where MLKAPS far outperforms the MKL. In this region, MLKAPS selects a good solution as expected, while \IntelOrthe  MKL uses a surprisingly inefficient tuning.  This corresponds to a blind spot in \IntelOrthe  MKL tuning strategy. We replicated this behavior on the same region on a Cascade Lake processor, but it is absent from the Saphire Rapids Processor.
Overall, with only $7,000$ samples on the KNM, MLKAPS matches or outperforms the MKL reference tuning on $74\%$ of the input space with a geomean speedup of $20\%$, and enabled us to identify a critical blind spot. 
\subsubsection{Results on the SPR}
\label{sec:dgetrf_experiment}

We repeated these experiments on the SPR for 7k, 15k, and 30k samples to show how MLKAPS performs when increasing the number of samples, as well as to highlight behavioral differences due to the hardware.

\begin{figure}[bt]
    \centering
    \includegraphics[width=1\linewidth]{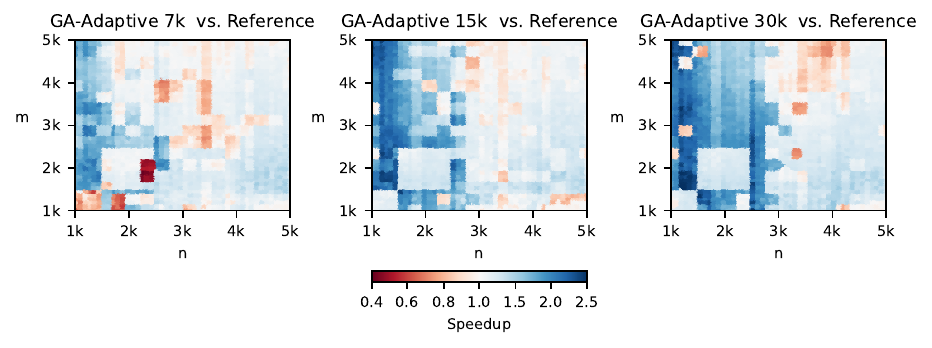}
    \caption{\textbf{(SPR): Speedup of MLKAPS's decision tree compared to the reference \IntelOrthe MKL configuration on \texttt{dgetrf} (LU)} for 7K, 15K, and 30K samples. 46x46 validation grid. The MLKAPS model performs better with more samples, with a geometric mean $\times 1.3$ speedup. The blocked patterns (Light-blue square region in the left-most plot) are due to the decision trees' partition strategy.}
    \label{fig:spr_mlkaps}
    \Description{This figure shows three 2d heatmaps of speedups of MLKAPS compared to the MKL reference, depending on the number of samples taken with MLKAPS. The 7k samples plot shows significant speedup regions, with some red regions for regressions. The 15k samples plot shows mostly speedup with very few regressions. The 30k samples plot shows higher speedups with more pronounced separation between regions, with more regressions in some regions. Overall, the 30k samples plot shows more speedups, meaning that the performance improve with the number of samples. A square region with limited speedup is present in all three plots.}
\end{figure}

Figure \ref{fig:spr_mlkaps} shows that the objective space is significantly different from the KNM processor and that MLKAPS can still find a significant speedup in modern architectures.
Similarly to the KNM experiments, there are clear regions of improvement and some regressions. As the knowledge of the model increases with the number of samples, we can observe that at 30K samples, almost no significant regression remains. With 30k samples, MLKAPS has a mean speedup of $30\%$. In addition to this mean speedup, we further distinguished the percentage of "regressions" points with a slowdown, and "progressions" with a speedup.
For 30k samples, we observe $15\%$ regressions with a mean slowdown of $\times0.10$, and $85\%$ of progressions with a mean speedup of $\times1.38$. 

\subsection{Comparison with other auto-tuners}
\subsubsection{Comparison to Optuna}
\label{sec:mlkaps_vs_optuna}
\begin{figure}[tb]
    \centering
    \vspace{-5mm}
    \begin{subfigure}{\linewidth}
        \centering
        \includegraphics{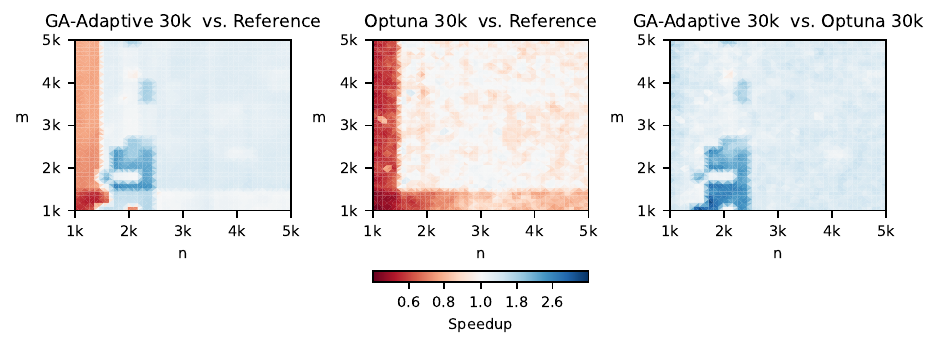}
    \end{subfigure}
    \caption{\textbf{(SPR): Speedup map of MLKAPS vs. Optuna on the MKL \texttt{dgeqrf} kernel.} MLKAPS has a geomean $18\%$ speedup over the MKL, but shows regions with significant regressions. In these regions, the MKL is near-optimal and very difficult to outperform.}
    \label{fig:optuna_vs_mlkaps}
\end{figure}

To show the benefits of transfer learning using MLKAPS' surrogate model, we compare our approach with Optuna~\cite{optuna_2019}, a renowned hyperparameter autotuner, on the \IntelOrthe MKL \texttt{dgeqrf} (QR) kernel. Optuna does not have a global model of the objective space, and the points are optimized individually.

This kernel shares the same two inputs and eight design parameters as \IntelOrthe MKL \texttt{dgetrf} (LU). We compare both tools on a $46\times46$ regular validation grid with 30k samples.

In Figure \ref{fig:optuna_vs_mlkaps} we observe that MLKAPS has a geomean speedup of $18\%$ on this kernel, improving the reference configuration on $85\%$ of the input space. The mean slowdown on regressions is $29\%$, the speedup on progressions $31\%$. This kernel has a better baseline configuration than \texttt{dgetrf}. Furthermore, MLKAPS outperforms Optuna on $98\%$ of the input space, with a geomean speedup of $36\%$.

These experiments show the benefits of using a surrogate model for transfer learning. However, we found that Optuna's ease of use and generalization capabilities were useful to tune MLKAPS algorithms, including the model hyperparameters to some extent. When tuning kernels for a single input, Optuna is competitive, but MLKAPS is more efficient when targeting large input spaces.

\subsubsection{Expert knowledge injection}

\begin{figure}[tb]
    \centering
    \includegraphics[width=\linewidth]{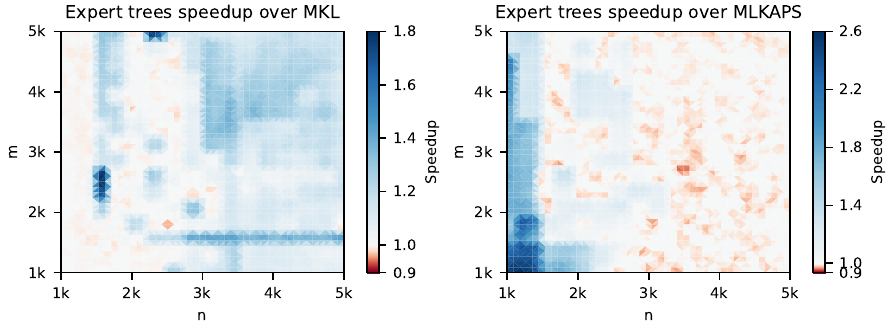}
    \caption{\textbf{(SPR): Expert tree combining MKL expert knowledge with MLKAPS auto-tuning solutions compared to MKL and MLKAPS decision trees.} MLKAPS 15k run. We eliminate all regressions (points remaining under 1.0 speedup are due to measurement noise) and bring significant improvements to the reference MKL configuration.  The geometric mean speedup over the MKL is $11\%$.}
    \label{fig:expert_trees_mlkaps_mkl}
    \Description{This figure shows two 2d speedup heatmaps. The first plot compares expert trees with the MKL reference. The plot shows mostly similar performance between the two methods, with regions with minor speedup between 1.2 and 1.4. The second plot compares expert trees with MLKAPS. Again, the plot shows similar performance between the two methods, except for the left and bottom bands, where the expert tree is significantly faster. The two speedup bands match the regions where MLKAPS was slower than the MKL in the previous figure.}
\end{figure}

The regressions observed in figure~\ref{fig:optuna_vs_mlkaps} are due to the MKL embedding an optimal tuning that is very hard to find using statistical sampling. In an industrial context, such regressions are not acceptable despite the achievement of speedups in other regions. It is worth noticing that each region being independent, it is possible to combine MLKAPS speedups with the MKL one in a single configuration tree to get the best of both worlds. This can be done by selecting the best solution on each input between the MKL reference and the MLKAPS optimized candidate. Such combinations ensure that we always select the best-known configuration, either from auto-tuning exploration or from previous expert knowledge. This method can also be applied with multiple MLKAPS runs to progressively refine the decision trees. Figure \ref{fig:expert_trees_mlkaps_mkl} shows the benefits of using this method: Expert trees outperform MKL for all inputs with a geometric mean speedup of $11\%$, and all regressions are removed. A 15k sample MLKAPS run was used to train the expert trees.

\subsubsection{Comparison to GPTune}
\label{sec:comparison_to_bayesian}

Bayesian optimization is a popular technique for auto-tuning hyperparameters in HPC~\cite{menon2020auto-tuning,combining_multitask_optimization}. GPTune~\cite{gptune} is a state-of-the-art black-box autotuning framework that leverages a sophisticated Bayesian optimization framework.
It builds a set of Gaussian processes and combines them using a linear model of coregionalization (LMC) to perform transfer learning and efficiently optimize multiple inputs at once.

GPTune and MLKAPS have different strategies for learning in a large design space. MLKAPS decouples the sampling and optimization phases by first building a surrogate model of the entire space. Once the surrogate model is trained, MLKAPS can optimize for any inputs. GPTune does not do this decoupling: The user must select the set of inputs for which GPTune will provide tuned design parameters, which also limits which inputs will be sampled. It can also perform random sampling for automated input selection.

The advantage of GPTune's approach is that the outputs are always validated against a directly measured sample; whereas MLKAPS' choices rely on the quality of the surrogate model. Recent versions of GPTune include the two-level transfer learning algorithm (TLA2), which allows the extrapolation of solutions on points that have not been explored, much like the MLKAPS decision tree(s).
However, GPTune extrapolation capabilities suffer from being constrained to sampling a limited set of inputs, therefore, completely missing performance cliffs, overfit to special cases,  or miss transitions in the objective space. This problem is mitigated in MLKAPS by using a surrogate model that does not restrict the input space as explained in Section~\ref{sec:optimizing}. Compensating for this would require GPTune to
increase the number of tasks, which we will see in a later section, does not scale.
Furthermore, for predicting configurations at runtime in a production environment, they would require their tool to be integrated with the application, while MLKAPS provides simple decision trees that can be embedded in the kernel.

We used the latest docker image on GPTune's repository (\texttt{liuyangzhuan/gptune:4.5}) with the LMC (Linear Coregionalization Model). Minor modifications were required to port the TLAII algorithm from the latest Github version, as it is not available in the Docker image. We compared both tools on two linear algebra kernels: the ScaLAPACK PDGEQRF (QR) described in their paper and the \IntelOrthe MKL \texttt{dgetrf} (LU).

\paragraph{ScaLAPACK PDGEQRF}

First, we ran MLKAPS on the ScaLAPACK \texttt{pdgeqrf} experiment described in the GPTune original paper~\cite{gptune}. This experiment was designed by the GPTune authors and optimizes a QR factorization kernel. They ran this experiment on up to 64 KNM nodes on the Cori cluster while we only have one KNM node available, which might explain differences in our results. In their paper, they optimized up to 10 input tasks with a total budget of 100 samples, which is significantly less than the usual budget required for MLKAPS experiments.

The current version of MLKAPS does not handle constrained optimization, so we reformulated the problem as a set of free parameters. The problem parameters are listed in Table \ref{tab:pdgeqrf_parameters}.
Original constraints are expressed as a set of inequalities (i.e., $mb \times p \times 8 \leq m$)
To satisfy the inequalities, we restricted three variables, $mb$, $nb$, and $n\_node$, and computed their upper and lower bounds according to the remaining variables. After this process, the admissible values for each bound variable lay in a convex interval, so we can replace the bound variable with a free parameter in $[0,1]$ that selects a point on the interval by linear interpolation (i.e. the previous example becomes $1 \leq mb \leq \frac{m}{8p}$). We ran GPTune on this transformed space to validate our approach and found that GPTune had a harder time fitting on these new variables. Note that only MLKAPS used this reformulation and that GPTune ran on the original problem.

\begin{table}
    \centering
    \begin{tabular}{|m{.15\textwidth}|m{.43\textwidth}|m{0.3\textwidth}|}
        \hline
        \cellcolor{gray!25} Name   & \cellcolor{gray!25} Description                                           & \cellcolor{gray!25} Reformulation \\ \hline
        $(m,n)$                    & Matrix size in the ($1^{st}$,$2^{nd}$) axis                               & Identical                         \\ \hline
        $p$                        & Process grid size along the $1^{st}$ axis                                 & Identical                         \\ \hline
        $mb\rightarrow\alpha$      & Block size along $m$                                                      &
        $mb = lerp(\alpha, 1, min(\frac{m}{8p}, 16))$                                                                                              \\ \hline
        $npernode\rightarrow\beta$ & Nb. of process per compute node                                           &
        $npernode = p + lerp(\beta, 0, 30 - p)$                                                                                                    \\ \hline
        $nb \rightarrow\gamma$     & Block size along $n$ ($np$ is the total number of processors, a constant) &
        $nb = lerp(\gamma, 1, min(\frac{np}{8npernode}, 16))$                                                                                      \\ \hline
    \end{tabular}
    \vspace{.5cm}
    \caption{\textbf{Parameters of PDGEQRF and their respective reformulation.} We note $lerp(\alpha, lb, ub)$ the linear interpolation between $lb$ and $ub$ with $\alpha \in [0, 1]$}
    \label{tab:pdgeqrf_parameters}
\end{table}

We choose 64 points chosen as a regular $8\times 8$ grid over matrix sizes $ 3072 \leq n,m \leq 8072$ as inputs for GPTune to fit their scalability limitation as described in a later experiment.

\begin{figure}
    \centering
    \includegraphics{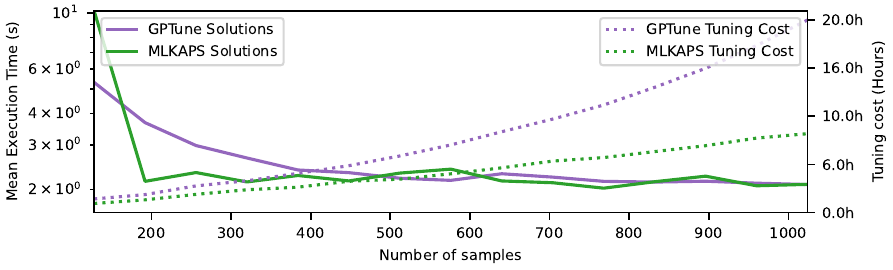}
    \caption{\textbf{(KNM): GPTune vs. MLKAPS on ScaLAPACK PDGEQRF (QR).} Both GPTune and MLKAPS converge to an equivalent optimized solution. Nevertheless, MLKAPS requires less tuning time and collected samples to find a near-optimal configuration.}
    \label{fig:gptune-scalapack}
    \Description{This plot is a lineplot with two axis. The left axis corresponds to the mean execution time in seconds. The bottom axis is the number of samples. We observe that MLKAPS converges to two seconds after only 200 samples. GPTune converges to two seconds after around 500 samples.
    The right axis shows the tuning cost in hours. MLKAPS collects 1000 samples in around 8 hours. GPTune collects 1000 samples in around 20 hours. Overall, MLKAPS is faster to run than GPTune, and converges faster.}
\end{figure}

We run GPTune and MLKAPS with an increasing number of samples up to 1024 and record the tuning cost and the performance achieved in figure~\ref{fig:gptune-scalapack}. Both tools converge to a mean execution time of $2.09s$ ($2.093$ for GPTune and $2.096$ for MLKAPS). Nevertheless, MLKAPS converges much faster than GPTune towards the optimal configuration: MLKAPS reaches optimal configurations with less than $200$ samples, while GPTune requires $500$ samples. We found that the objective in this experiment is almost entirely dominated by the parameter $p$, which explains why such a low number of samples is required for convergence.

Moreover, MLKAPS has a lower tuning cost than GPTune for all sample counts and is up to $2.44\times$ faster for 1024 samples. This leads to one of the main pitfalls of GPTune's approach compared to MLKAPS: its optimization pipeline does not scale to a large number of samples and grid points.

\paragraph{MKL \texttt{dgetrf}}

\begin{figure}
    \centering
    \includegraphics{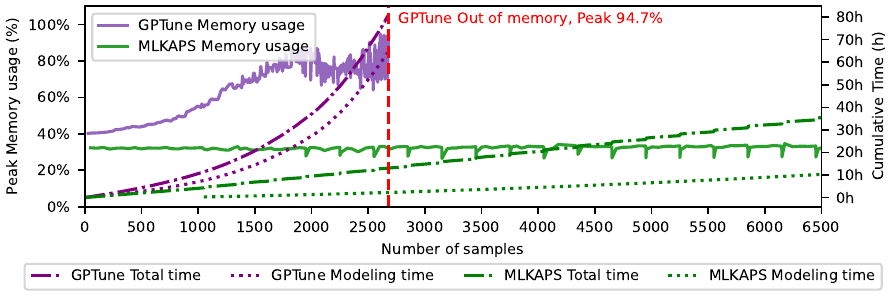}
    \caption{\textbf{(KNM): Evolution of peak memory usage and sampling time on the \texttt{dgetrf} (LU) experiment, with 16 tasks and a 7k sampling budget.} GPTune memory and modeling time increase non-linearly with the number of samples. After 2512 samples, GPTune was killed by the operating system, having consumed all available memory. The instability in the GPTune memory usage after 1500 samples is likely due to the operating system swapping memory regions to disk.
    MLKAPS exhibits linear scaling in time and constant memory usage, handling a large number of samples well.
}
    \label{fig:gptune_memory_dgetrf}
    \Description{This figure is a line plot with two axis. On the left is the Peak memory usage. The bottom axis is the the number of samples. We observe that GPTune memory usage quickly grows to 80\%, becomes unstable, and stops at 94.7\% for 2512 samples due to an out of memory error. MLKAPS is around 35\% memory usage and stays constant. The right axis shows the cumulative time in hours. The GPTune total execution time grows non-linearly up to 80 hours for 2512 samples. A second line shows the GPTune modeling time, which demonstrates a similar growth and closely follows the total time. This means that the modeling dominates the total time. MLKAPS total time ends at around 35 hours after 7000 samples. The modeling time sits at around 10 hours for 7000 samples. Overall, GPTune scales non-linearly in memory and time, crashes with 2500 samples, while MLKAPS scales linearly in memory and time, and successfully completes the target 7000 samples.}
\end{figure}

To confirm GPTune scalability issues, we decided to run the tool on \IntelOrthe \texttt{dgetrf} (LU) kernel presented earlier in Section~\ref{sec:dgetrf_experiment}. We previously saw that MLKAPS continuously improved with an increasing number of samples up to 30k, leading us to believe that this experiment is significantly harder than ScaLAPACK \texttt{pdgeqrf} as it requires more samples.

For this experiment, we run GPTune with $8$ random input tasks and then use TLAII for predicting on a $46\times 46$ grid. The run crashed due to an out-of-memory issue, as visible in figure~\ref{fig:gptune_memory_dgetrf} which represents memory usage according to the number of samples. We found that GPTune's approach suffers from a scalability issue, both in memory and execution time, and is dependent on the number of samples and the number of tasks. From our understanding, this issue comes from the covariance matrix needed by LCM of size $\epsilon\delta \times \epsilon\delta $ (Sec. 3.3 in~\cite{gptune}) with $\epsilon$ the number of samples per task, and $\delta$ the number of tasks. MLKAPS scales linearly in memory and time with the number of samples, and most of the runtime is spent collecting samples. Overall, MLKAPS achieves a $15\%$ speedup on \texttt{dgetrf} when compared to GPTune on the KNM and finishes in half the time with $\times 3$ the number of samples. Overall, we believe that both MLKAPS and GPTune approaches are valid, but target different scopes:
\begin{itemize}
    \item MLKAPS is most useful for building decision trees for predicting configurations on a large range of inputs, in cases where thousands of measures are feasible.
    \item On the opposite side, GPTune provides a sophisticated optimization framework that has the advantage of handling constraints over the parameter space and is well suited to optimize thoroughly over a few well-chosen inputs but cannot scale to larger experiments. 
\end{itemize}

\section{Discussions and Future Works}
\label{sec:limitations}

Our design already provides very encouraging results. Future works to improve MLKAPS include:
\begin{itemize}
    \item Extend our use-cases to more kernels and hardware architectures.
    \item Handle constraint without explicit reformulation from the user.
    \item Improve the decision trees, measure both overhead, interpretability, and performance. Currently, MLKAPS uses one tree per design parameter, and we are considering merging into a single and using more complex splitting criteria.
    \item Dynamic Optimization Grid: Currently, MLKAPS uses a fixed regular grid to optimize on. However, this method was shown to miss local optima in some cases and is inefficient as neighboring points in the grid often share similar configurations, which in turn induces useless splits in the tree(s).
    \item Extending support for expert-knowledge injection to leverage existing tunings in cases similar to \texttt{dgetrf}, or allow user to inject better solution in MLKAPS model enabling to re-tune the solution provided by the black-box model approach.
    \item Introducing a metric for the surrogate model that measures its optimization capabilities.
    \item Our preliminary works show that Optuna~\cite{optuna_2019} may be used to tune MLKAPS hyper-parameters, relieving the non-expert end-user from complex understanding and model tuning  process.
\end{itemize}

\section{Conclusion} \label{sec:conclusion}

This article shows how manual tuning is
error-prone due to human bias and often relies on inefficient
optimization methods. We introduced MLKAPS, an auto-tuning tool that generates decision trees that can be embedded into HPC kernels.  We described the implementation of GA-Adaptive, an optima-biased sampler that outperformed the other sampling strategies.

We applied MLKAPS to an industrial parallelized kernel taken from \IntelOrthe MKL and compared
how the different sampling techniques led to different surrogates and tuning
performances. GA-Adaptive offers better tuning results than other techniques and improves the surrogate's local accuracy. We compared the tuning solutions
proposed by MLKAPS with the reference hand-tuned \IntelOrthe MKL on the \texttt{dgetrf} kernel. We showed that MLKAPS matches or outperforms \IntelOrthe \texttt{MKL} on $85\%$ of the input space with an average speedup of $38\%$.

We compared MLKAPS to the state-of-the-art auto-tuner GPTune and Optuna on kernels from both ScaLAPACK and \IntelOrthe MKL. MLKAPS is significantly more scalable and outperforms GPTune on problems with large design space. MLKAPS global GBDT model allows it to outperform Optuna for the same number of samples.

This article highlighted the advantages of an auto-tuning tool capable of generating
decision trees for dynamic runtime tuning. MLKAPS is open-source and available at \href{https://github.com/MLCGO/MLKAPS}{https://github.com/MLCGO/MLKAPS}.

{\footnotesize This project received funding from an Intel Higher Education grant.}

\bibliographystyle{ACM-Reference-Format}
\bibliography{bibliography}
\end{document}